# Cooperative Diversity With Mobile Nodes: Capacity Outage Rate and Duration

Nikola Zlatanov, *Student Member, IEEE*, Zoran Hadzi-Velkov, *Senior Member, IEEE*, George K. Karagiannidis, *Senior Member, IEEE*, and Robert Schober, *Fellow, IEEE*

*Abstract*—The outage probability is an important performance measure for cooperative diversity schemes. However, in mobile environments, the outage probability does not completely describe the behavior of cooperative diversity schemes since the mobility of the involved nodes introduces variations in the channel gains. As a result, the capacity outage events are correlated in time and second-order statistical parameters of the achievable information-theoretic capacity such as the average capacity outage rate (AOR) and the average capacity outage duration (AOD) are required to obtain a more complete description of the properties of cooperative diversity protocols. In this paper, assuming slow Rayleigh fading, we derive exact expressions for the AOR and the AOD of three well-known cooperative diversity protocols: variable-gain amplify-and-forward, decode-and-forward, and selection decode-and-forward relaying. Furthermore, we develop asymptotically tight high signal-to-noise ratio (SNR) approximations, which offer important insights into the influence of various system and channel parameters on the AOR and the AOD. In particular, we show that on a double-logarithmic scale, similar to the outage probability, the AOR asymptotically decays with the SNR with a slope that depends on the diversity gain of the cooperative protocol, whereas the AOD asymptotically decays with a slope of $-1/2$ independent of the diversity gain.

*Index Terms*—Average outage duration (AOD), average outage rate (AOR), cooperative diversity, Doppler effect, Rayleigh fading.

## I. INTRODUCTION

THE capacity outage probability (OP) is an important performance measure in wireless communication systems with delay constraints operating over slow fading channels [1], [2]. The OP is the probability that the channel capacity is smaller than a given transmission rate [3], [4]. Cooperative diversity is an efficient means to improve the OP of wireless systems by exploiting spatially distributed nodes (also referred to as relays) to effectively synthesize a virtual array that emulates the operation of a multiantenna transceiver [5]–[19].

In particular, Laneman *et al.* developed in [9] several simple repetition-based cooperative protocols: amplify-and-forward relaying (AF), decode-and-forward relaying (DF), selection decode-and-forward relaying (SR), and incremental relaying. The performance of these protocols was characterized in terms of the asymptotic OP for high signal-to-noise ratios (SNRs), and it was shown analytically that, except for DF relaying, these protocols achieve full diversity.

In systems with mobile nodes, the channel gains are time varying. In this case, for the OP to be a relevant performance measure, two conditions have to be fulfilled [20]. 1) The delay constraint has to be small compared to the channel coherence time [2]. In other words, for the duration of one codeword, the channel has to be practically constant. 2) The codewords have to be sufficiently long such that capacity-approaching codes exist. Both 1) and 2) involve approximations which, at the expense of an increase in bandwidth, can be made arbitrarily tight by shortening the coding block and increasing the number of symbols per coding block, respectively. Nevertheless, the channel gain and the corresponding instantaneous channel capacity will change slowly from one coding block to the next. As a result, the channel capacities in neighboring coding blocks are correlated in time. Thus, capacity outage events are correlated in time as well. This correlation is not reflected in the OP itself but is captured by the average capacity outage rate (AOR) and the average capacity outage duration (AOD). The notions of AOR and AOD have initially been introduced for opportunistic relaying systems in [21]. A similar definition of second-order outage statistics was also used for multiple-input–multiple-output (MIMO) systems in [22].

The AOR and AOD provide important information for the design of wireless communication systems. For example, in systems with automatic repeat request (ARQ), the waiting time before a packet is retransmitted should be chosen larger than the AOD to avoid unsuccessful retransmissions. In multiuser systems, a scheduling slot typically comprises several coding blocks. Making the size of a scheduling slot larger than the AOD will guarantee a low number of unsuccessful scheduling attempts of users who were in outage in the previous scheduling slot. Furthermore, for systems with stringent energy constraints, the transmitters and receivers can be switched off for at least the AOD to conserve energy if a packet cannot be decoded successfully, as it is unlikely that the following packets would be decoded successfully. In this case, the AOR indicates the frequency with which the receivers are switched on and off. While these examples illustrate the usefulness of the AOR and AOD for the design of general wireless networks, their importance is further enhanced in cooperative diversity systems. In particular, the involvement of multiple network nodes increases the overhead associated with unsuccessful retransmissions and scheduling attempts as well as the amount of energy consumed for unsuccessful decoding attempts.

In this paper, we present an analytical framework for calculation of the AOR and AOD of cooperative diversity systems with mobile nodes. Specifically, we derive exact expressions for both parameters for cooperative diversity systems employing the aforementioned AF, DF, and SR protocols, respectively, which are assumed to operate over slowly time-varying Rayleigh fading channels. We also derive simple closed-form high SNR approximations for the AOR and AOD. These approximations provide significant insight into the dependence of the AOR and AOD on various system and channel parameters such as the Doppler frequency and the data rate. This insight facilitates the design of cooperative diversity systems and the comparison of different protocols.

The remainder of this paper is organized as follows. In Section II, we present the system and channel models. The AOR and AOD are defined in Section III, and exact analytical expressions for the AF, DF, and SR protocols are provided. In Section IV, we derive the respective asymptotic high SNR approximations for the AOR and AOD. In Section V, the AOR and AOD of the considered cooperative diversity protocols are compared based on numerical results. Conclusions are drawn in Section VI.

## II. COOPERATIVE DIVERSITY SYSTEMS WITH MOBILE NODES

In this section, we present the system model and the adopted slow Rayleigh fading channel model.

### A. System Model

We consider the same cooperation scenario as in [9], where a given source–destination pair communicates over a relay $R$ by utilizing one of the three considered half-duplex protocols: AF, DF, and SR. The source $S$ and the destination $D$ communicate over two orthogonal subchannels: the direct subchannel ($S \to D$ link) and the relayed subchannel, which consists of the $S \to R$ link and the $R \to D$ link. Orthogonality of the subchannels is achieved through a suitable orthogonal multiplexing scheme such as time-division, frequency-division, or code-division multiplexing.

Let us denote the channel gains of the $S \to D$, $S \to R$, and $R \to D$ links by $X(t)$, $Y(t)$, and $Z(t)$, respectively, with average squared channel gains $E[X^2(t)] = \Omega_{SD} = \Omega_X$, $E[Y^2(t)] = \Omega_{SR} = \Omega_Y$, and $E[Z^2(t)] = \Omega_{RD} = \Omega_Z$, respectively. Here, $E[\cdot]$ denotes statistical expectation. Since the average squared channel gains can be arbitrarily chosen, we assume without loss of generality that $S$ and $R$ transmit with equal powers $P_T$. All nodes are impaired by additive white Gaussian noise with single-sided power spectral density $N_0$.

For each of the three considered protocols, $S$ broadcasts the information-bearing signal, and both $D$ and $R$ receive it. If the AF protocol is utilized, $R$ amplifies the received signal (along with its own noise) and forwards it to $D$ over the $R \to D$ link, and $D$ combines the replicas received over the $S \to D$ and $R \to D$ links in an attempt to decode. In case of the DF protocol,

$R$ receives the information-bearing signal, attempts to decode the received message, and then re-encodes and retransmits the estimated message over the $R \to D$ link. For the SR protocol, if the channel gain $Y(t)$ falls below a certain threshold $Y_0$, $S$ simply retransmits the same packet over the $S \to D$ link while $R$ remains silent; otherwise $R$ re-encodes and retransmits the estimated message over the $R \to D$ link.

The transmissions from $S$ are organized into coding blocks of duration $T$, where each coding block is occupied by one codeword. We assume that the number of symbols per coding block is sufficiently large such that capacity-approaching codes exist.

### B. Channel Model and Mobility of the Nodes

We assume there is no line-of-sight between any of the involved nodes and all links are affected by mutually independent Rayleigh fading. Thus, at time $t$, the three channel gains follow the Rayleigh probability density function (pdf)

$$f_\alpha(x) = \frac{2x}{\Omega_\alpha} \exp\left(-\frac{x^2}{\Omega_\alpha}\right), \qquad x \geq 0, \qquad \alpha \in \{X, Y, Z\} \tag{1}$$

with cumulative distribution function (cdf)

$$F_\alpha(x) = \Pr\{\alpha \leq x\} = 1 - \exp\left(-\frac{x^2}{\Omega_\alpha}\right), \qquad x \geq 0, \\ \alpha \in \{X, Y, Z\}. \tag{2}$$

Furthermore, due to the mobility of the nodes, the channel gains are time varying. Here, we assume that the considered application has severe delay constraints such that the coding blocks are short compared to the coherence time $T_{\text{coh}}$ of the channel, i.e., $T \ll T_{\text{coh}}$. Thus, we can assume that the channel gains $X(t)$, $Y(t)$, and $Z(t)$ are practically constant for the duration of one coding block but change slowly from one coding block to the next.

As usual, we model the channel gains $X(t)$, $Y(t)$, and $Z(t)$ as time-correlated random processes. The degree of variability (and the coherence time) of the channels depends on the power spectral density (a.k.a. "Doppler spectrum") of the channel gains, which is determined by the scattering environment and the mobility of the involved transmitters and receivers. Here, we consider a 2-D isotropic scattering environment and mobility of $S$, $R$, and $D$, such that the $S \to D$, $S \to R$, and $R \to D$ links can be modeled as independent mobile-to-mobile Rayleigh fading channels. Such channels have been extensively studied in [23] and [24] and the autocovariance function and Doppler spectrum of the channel gains are specified in [23, eq. (35)] and [23, eq. (41)], respectively. It is worth noting that when the transmitter or the receiver is static, the mobile-to-mobile channel model simplifies to the "classical" Jake's fading channel model [25], [26].

In the following, we will make use of the fact that the time derivatives of the gains of the mobile-to-mobile channels $\dot{X}(t)$, $\dot{Y}(t)$, and $\dot{Z}(t)$ are independent from the respective gains $X(t)$, $Y(t)$, and $Z(t)$ themselves and follow a zero-mean Gaussian pdf [24], [25]. The variance of the derivatives of the channel gains is given by [24]

$$\sigma_{\dot{\alpha}}^2 = \pi^2 \Omega_\alpha f_{m,\alpha}^2, \qquad \alpha \in \{X, Y, Z\} \tag{3}$$

where $f_{mX} = \sqrt{f_{mS}^2 + f_{mD}^2}$, $f_{mY} = \sqrt{f_{mS}^2 + f_{mR}^2}$, and $f_{mZ} = \sqrt{f_{mR}^2 + f_{mD}^2}$. Here, $f_{mS}$, $f_{mR}$, and $f_{mD}$ denote the maximum Doppler rates introduced by the mobility of $S$, $R$, and $D$, respectively. Since the coherence time of the channel is reciprocal to the maximum Doppler frequency [20], the condition $T \ll T_{\text{coh}}$ implies

$$f_{mS}T \ll 1 \quad f_{mR}T \ll 1 \quad f_{mD}T \ll 1. \quad (4)$$

For a more detailed discussion of the mobile-to-mobile channel model we refer the interested reader to [23] and [24].

### III. CAPACITY OUTAGE RATE AND DURATION

In this section, we derive exact expressions for the AOR and AOD for the considered cooperative diversity protocols. However, first, we develop general formulas for the AOR and AOD.

#### A. Derivation of the AOR and AOD

At a given time $t$, the instantaneous mutual information between the signals at the input and the output of the considered cooperative diversity systems with equivalent end-to-end fading channel $G(t)$ is given by

$$I(t) = \frac{1}{2} \log_2 \left[ 1 + \Gamma_0 G^2(t) \right] \quad (5)$$

where $\Gamma_0$ denotes the *transmit SNR* (also referred to as the SNR without fading), defined as $\Gamma_0 = P_T/N_0$.

A capacity outage event occurs when the mutual information $I(t)$ drops below some fixed target spectral efficiency $R_0$ [20], $I(t) \leq R_0$, or, equivalently, if

$$G(t) \leq G_0 \quad (6)$$

where $G_0$ is the *outage threshold*, given by $G_0 = \sqrt{(2^{2R_0} - 1)/\Gamma_0}$. Thus, the occurrence of a "deep fade" in the equivalent end-to-end channel $G(t)$ yields a capacity outage event of the cooperative system, because this channel cannot support reliable communication between $S$ and $D$ at the desired information rate $R_0$.

Based on (6), the OP of the cooperative diversity system equals the cdf of $G(t)$ evaluated at the outage threshold $G_0$

$$P_{\text{out}} = \Pr\{I(t) \leq R_0\} = F_G(G_0). \quad (7)$$

The asymptotic OPs for the considered AF, DF, and SR protocols have been determined in [9] for high transmit SNR, whereas, in this paper, as a byproduct of the derivation of the AOD, we provide the respective exact expressions valid for arbitrary transmit SNR. In this context, we emphasize again that for (7) to be a meaningful performance measure, each coding block has to accommodate a sufficiently large number of symbols such that capacity-approaching codes exist, and each coding block has to be sufficiently short such that $G(t)$ is approximately constant over the entire codeword. Since the equivalent channel gain $G(t)$ is a protocol-dependent function of the channel gains

$$G(t) = f[X(t), Y(t), Z(t)] \quad (8)$$

and $X(t)$, $Y(t)$, and $Z(t)$ satisfy (4), $G(t)$ is also a slowly time-varying random process with negligible variability during each coding block. Thus, an outage event (6) in the equivalent end-to-end channel affects at least one coding block. However, an outage event can also affect several consecutive coding blocks. Thus, in general, a capacity outage event lasts $kT$ (seconds), where $k \in \{1, 2, 3, \ldots\}$.

Let us consider a time interval of duration $W$, which spans many channel coherence times, i.e., $W \gg T_{\text{coh}}$, and assume that $L$ capacity outage events occur during this interval. The AOR is defined as the occurrence rate of the capacity outage events during the considered interval, i.e.,

$$N_I \stackrel{\text{def}}{=} \frac{L}{W}. \quad (9)$$

Considering (5) and assuming $G(t)$ is a stationary random process, (9) can be computed using Rice's formula [25], [26, eq. (2.101)]. In other words, the AOR can be estimated from the level crossing rate (LCR) of random process $G(t)$ evaluated at $G_0$, yielding

$$N_I = N_G(G_0) = \int_0^\infty \dot{g} f_{G\dot{G}}(G_0, \dot{g}) d\dot{g} \quad (10)$$

where $\dot{G}$ denotes the time derivative of random process $G$, and $f_{G\dot{G}}(g, \dot{g})$ is the joint pdf of $G$ and $\dot{G}$.[1]

Furthermore, let us denote the respective durations of the $L$ capacity outage events in the considered time interval of duration $W$ by $k_1 T, k_2 T, \ldots$, and $k_L T$. Since the AOD is by definition the average duration of the observed capacity outage events, we have

$$T_I \stackrel{\text{def}}{=} \frac{k_1 T + k_2 T + \cdots + k_L T}{L}$$
$$= \frac{(k_1 T + k_2 T + \cdots + k_L T)/W}{L/W}. \quad (11)$$

Since the numerator and the denominator of the right-hand side of (11) are the OP and the AOR, respectively, we obtain for the AOD

$$T_I = \frac{P_{\text{out}}}{N_I} = \frac{F_G(G_0)}{N_G(G_0)}. \quad (12)$$

Thus, once the OP and the AOR are computed from (7) and (10), respectively, the AOD can be determined from (12). In the remainder of this section, we exploit these relations to compute the AOR and the AOD of AF, DF, and SR relaying. However, first, we briefly consider direct transmission to establish a reference for the considered cooperative diversity schemes.

#### B. Direct Transmission

For direct transmission between $S$ and $D$ the maximum average mutual information is given by [9, eq. (10)]

$$I_D(t) = \log_2 \left[ 1 + \Gamma_0 X^2(t) \right]. \quad (13)$$

---
[1]We drop the time index $t$, whenever this is possible without causing ambiguity.

A capacity outage event occurs when the mutual information $I_D$ drops below the target spectral efficiency $R_0$, or, equivalently, if

$$X(t) \leq X_0 \tag{14}$$

with the outage threshold $X_0 = \sqrt{(2^{R_0} - 1)/\Gamma_0}$. Thus, the OP for direct transmission is given by

$$P_{\text{out}} = \Pr\{I_D \leq R_0\} = F_X(X_0) = 1 - \exp\left(-\frac{X_0^2}{\Omega_X}\right). \tag{15}$$

Furthermore, based on (10), the AOR $N_I$ is obtained as

$$N_I(R_0) = N_X(X_0) = \sqrt{\frac{2\sigma_{\dot{X}}^2}{\pi}} \frac{X_0}{\Omega_X} \exp\left(-\frac{X_0^2}{\Omega_X}\right). \tag{16}$$

The AOD $T_I$ is obtained by inserting (15) and (16) into (12).

### C. Variable-Gain AF Relaying

Variable-gain AF relays set the amplification gain to $\sqrt{1/(Y^2(t) + C_1)}$, where $C_1 = 1/\Gamma_0$, in order to fix the power of the retransmitted signal to $P_T$. Thus, based on [9], (12), and (13), $G(t)$ is given by

$$G(t) = \sqrt{X^2(t) + \frac{Y^2(t) Z^2(t)}{Y^2(t) + Z^2(t) + C_1}}. \tag{17}$$

*Theorem 1:* The OP of a cooperative diversity system utilizing variable-gain AF relaying is given by

$$F_G(G_0) = \int_0^{G_0^2} \frac{1}{\Omega_X} \exp\left(-\frac{G_0^2 - a}{\Omega_X}\right)$$
$$\times \left[1 - 2\sqrt{\frac{a(a+C_1)}{\Omega_Z \Omega_Y}}\right.$$
$$\left. \times \exp\left(-\frac{\Omega_Y + \Omega_Z}{\Omega_Y \Omega_Z} a\right) K_1\left(2\sqrt{\frac{a(a+C_1)}{\Omega_Z \Omega_Y}}\right)\right] da \tag{18}$$

where $K_1(\cdot)$ is the modified first-order Bessel function of the second kind [27, eq. (9.6.2)].

*Proof:* See Appendix A.

Note that the integral in (18) can be efficiently and accurately evaluated by applying the Gauss–Legendre numerical quadrature rule [27, eq. (25.4.29)–(25.4.30)].

*Theorem 2:* The AOR of a cooperative diversity system utilizing variable-gain AF relaying is given by

$$N_I(R_0) = N_G(G_0) = \sqrt{\frac{2}{\pi}} \frac{1}{\Omega_X \Omega_Y \Omega_Z} \exp\left(-\frac{G_0^2}{\Omega_X}\right)$$
$$\times \int_0^{G_0^2} da \int_0^\infty \left[(G_0^2 - a)\sigma_{\dot{X}}^2 + \frac{a^2 t^3 (a + C_1)^2}{(at+1)(at + C_1 t + 1)^2} \sigma_{\dot{Y}}^2\right.$$
$$\left. + \frac{a}{(at+1)^2 (at + C_1 t + 1)} \sigma_{\dot{Z}}^2\right]^{1/2}$$
$$\times \frac{(at+1)(at + C_1 t + 1)}{t^2} \exp\left[-a\left(\frac{1}{\Omega_Y} + \frac{1}{\Omega_Z} - \frac{1}{\Omega_X}\right)\right.$$
$$\left. - \left(\frac{at(a + C_1)}{\Omega_Z} + \frac{1}{t \Omega_Y}\right)\right] dt. \tag{19}$$

*Proof:* See Appendix A.

The double integral in (19) can be evaluated efficiently and accurately by a product of two quadrature rules: a Gauss–Legendre rule [27, eq. (25.4.29)–(25.4.30)] for integration over variable $a$ and a Gauss–Laguerre rule [27, eq. (25.4.45)] for integration over variable $t$.

The AOD $T_I$ of variable-gain AF relaying is obtained by inserting (18) and (19) into (12).

### D. DF Relaying

When DF relaying is considered, an exact expression for the maximum average mutual information can be obtained only under the assumption of repetition coding and full decoding of the source message by the relay [9], [11, eq. (15)]. In this case, $G(t)$ is given by

$$G(t) = \min\{Y(t), U(t)\} \tag{20}$$

where $U(t)$ is an auxiliary random process, defined as

$$U(t) = \sqrt{X^2(t) + Z^2(t)}. \tag{21}$$

The OP of a cooperative diversity system employing DF relaying is given by

$$F_G(G_0) = 1 - \Pr\{Y > G_0\} \Pr\{U > G_0\} \tag{22}$$

where [29]

$$\Pr\{U > G_0\}$$
$$= \begin{cases} \frac{\Omega_X}{\Omega_X - \Omega_Z} e^{-G_0^2/\Omega_X} + \frac{\Omega_Z}{\Omega_Z - \Omega_X} e^{-G_0^2/\Omega_Z}, & \Omega_X \neq \Omega_Z \\ \exp(-G_0^2/\Omega_X)(1 + G_0^2/\Omega_X), & \Omega_X = \Omega_Z \end{cases} \tag{23}$$

and

$$\Pr\{Y > G_0\} = \exp\left(-\frac{G_0^2}{\Omega_Y}\right). \tag{24}$$

From the time derivative of both sides of (20)

$$\dot{G} = \begin{cases} \dot{Y}, & Y \leq U \\ \dot{U}, & Y > U \end{cases} \tag{25}$$

and the independence of the channel gains $Y$ and $U$, the joint pdf of $G$ and $\dot{G}$ is determined as

$$f_{G\dot{G}}(g, \dot{g}) = f_{Y\dot{Y}}(g, \dot{g}) \Pr\{U > g\} + f_{U\dot{U}}(g, \dot{g}) \Pr\{Y > g\}. \tag{26}$$

Applying (26) in (10) yields the following expression for the AOR:

$$N_I(R_0) = N_G(G_0)$$
$$= N_Y(G_0) \Pr\{U > G_0\} + N_U(G_0) \Pr\{Y > G_0\} \tag{27}$$

where $\Pr\{U > G_0\}$ and $\Pr\{Y > G_0\}$ are given by (23) and (24), respectively. In (27), $N_Y(G_0)$ is the LCR of Rayleigh random process $Y(t)$, which is well known and given by [24, eq. (2)]

$$N_Y(G_0) = \sqrt{\frac{2\sigma_{\dot{Y}}^2}{\pi}} \frac{G_0}{\Omega_Y} \exp\left(-\frac{G_0^2}{\Omega_Y}\right) \quad (28)$$

whereas $N_U(G_0)$ is the LCR of random process $U(t)$, which is given in the following theorem.

*Theorem 3:* The LCR of random process $U(t)$ is given by

$$N_U(G_0) = \sqrt{\frac{2}{\pi}} \frac{\sigma_{\dot{X}}^3}{\sigma_{\dot{Z}}^2 - \sigma_{\dot{X}}^2} \frac{\exp\left(-G_0^2/\Omega_X\right)}{\Omega_X \Omega_Z} \left(\frac{G_0}{\sqrt{W(G_0)}}\right)^3$$
$$\times \exp\left(W(G_0)\right) \left[\Gamma\left(\frac{3}{2}, W(G_0)\right) - \Gamma\left(\frac{3}{2}, \frac{\sigma_{\dot{Z}}^2}{\sigma_{\dot{X}}^2} W(G_0)\right)\right]$$
(29)

where $\Gamma(\cdot,\cdot)$ is the incomplete Gamma function, defined in [27, eq. (6.5.3)], and $W(\cdot)$ is defined as

$$W(G_0) = \frac{G_0^2(\Omega_X - \Omega_Z)}{\Omega_X \Omega_Z} \frac{\sigma_{\dot{X}}^2}{\sigma_{\dot{Z}}^2 - \sigma_{\dot{X}}^2}. \quad (30)$$

*Proof:* See Appendix B.

When $\Omega_X = \Omega_Z$, according to Appendix B, (29) simplifies to

$$N_U(G_0) = \frac{4G_0^3}{3\sqrt{2\pi}} \frac{\exp\left(-G_0^2/\Omega_X\right)}{(\Omega_X)^2} \frac{\sigma_{\dot{Z}}^3 - \sigma_{\dot{X}}^3}{\sigma_{\dot{Z}}^2 - \sigma_{\dot{X}}^2} \quad (31)$$

where $\sigma_{\dot{X}}^2$ and $\sigma_{\dot{Z}}^2$ are given by (3).

When $\Omega_X = \Omega_Z$ and $f_{mS} = f_{mR} = f_{mD}$ then $\sigma_{\dot{X}} = \sigma_{\dot{Z}}$, and the limit operation $\sigma_{\dot{X}} \to \sigma_{\dot{Z}}$ can be applied in (31) to obtain

$$N_U(G_0) = \frac{\sqrt{2}\sigma_{\dot{X}} G_0^3}{\sqrt{\pi}} \frac{\exp\left(-G_0^2/\Omega_X\right)}{(\Omega_X)^2}. \quad (32)$$

A closed-form expression for the AOR $N_I$ of the DF protocol is obtained by inserting (23), (24), (28), and (29) [or (31) or (32)] into (27). Furthermore, the AOD is obtained by inserting (22) and (27) into (12).

*E. Selection DF Relaying*

The selection DF relaying protocol activates the relay only if the measured channel gain $Y(t)$ is above a given threshold $Y_0$ such that the received codeword can be successfully decoded. If the relay is activated, it decodes the message and forwards it over the $R \to D$ link to the destination; otherwise $S$ retransmits the message. Thus, based on [9, eq. (19)], for the case of repetition coding at the relay, $G(t)$ is obtained as

$$G(t) = \begin{cases} \sqrt{2}X(t), & Y(t) \leq Y_0 \\ U(t), & Y(t) > Y_0 \end{cases} \quad (33)$$

where $U(t)$ is defined in (21).

The OP of a cooperative diversity system utilizing SR relaying can be expressed as

$$F_G(G_0) = \Pr\left\{\sqrt{2}X \leq G_0\right\} \Pr\{Y \leq Y_0\}$$
$$+ \Pr\{U \leq G_0\} \Pr\{Y > Y_0\} \quad (34)$$

where

$$\Pr\{Y \leq Y_0\} = 1 - \Pr\{Y > Y_0\} = 1 - \exp\left(-\frac{Y_0^2}{\Omega_Y}\right) \quad (35)$$

$$\Pr\left\{\sqrt{2}X \leq G_0\right\} = 1 - \exp\left(-\frac{G_0^2}{2\Omega_X}\right) \quad (36)$$

and $\Pr\{U \leq G_0\} = 1 - \Pr\{U > G_0\}$ is calculated from (23).

By carefully analyzing the conditions for downward crossings of random process $G(t)$ of the outage threshold $G_0$, the following four independent downward crossing events can be identified:

i) downward crossing of $G(t) = \sqrt{2}X(t)$, if $Y(t) \leq Y_0$;
ii) downward crossing of $G(t) = U(t)$, if $Y(t) > Y_0$;
iii) downward crossing of $G(t)$ when $Y(t)$ switches from condition $Y(t) \leq Y_0$ to condition $Y(t) > Y_0$, but only if $\sqrt{2}X(t) > G_0$ and $U(t) < G_0$;
iv) downward crossing of $G(t)$ when $Y(t)$ switches from condition $Y(t) > Y_0$ to condition $Y(t) \leq Y_0$, but only if $U(t) > G_0$ and $\sqrt{2}X(t) < G_0$.

Thereby, event i) occurs with probability $\Pr\{Y \leq Y_0\}$, event ii) occurs with probability $\Pr\{Y > Y_0\}$, event iii) occurs with probability $\Pr\{\sqrt{2}X > G_0 \cap U < G_0\}$, and event iv) occurs with probability $\Pr\{\sqrt{2}X < G_0 \cap U > G_0\}$. Thus, the AOR of a cooperative diversity system employing SR relaying is given by

$$N_I(R_0) = N_G(G_0)$$
$$= N_{\sqrt{2}X}(G_0) \Pr\{Y \leq Y_0\} + N_U(G_0) \Pr\{Y > Y_0\}$$
$$+ N_Y(Y_0) \Pr\left\{\sqrt{2}X(t) > G_0 \cap U(t) < G_0\right\}$$
$$+ N_Y(Y_0) \Pr\left\{\sqrt{2}X(t) < G_0 \cap U(t) > G_0\right\}.$$
(37)

Using (16), $N_{\sqrt{2}X}(G_0)$ is obtained as

$$N_{\sqrt{2}X}(G_0) = N_X\left(\frac{G_0}{\sqrt{2}}\right) = \sqrt{\frac{\sigma_{\dot{X}}^2}{\pi}} \frac{G_0}{\Omega_X} \exp\left(-\frac{G_0^2}{2\Omega_X}\right). \quad (38)$$

Furthermore, it can be shown that

$$\Pr\left\{\sqrt{2}X > G_0 \cap U < G_0\right\}$$
$$= \begin{cases} e^{-G_0^2/(2\Omega_X)} - e^{-G_0^2/(2\Omega_X)}e^{-G_0^2/(2\Omega_Z)} \\ + \dfrac{\Omega_X}{\Omega_Z - \Omega_X}\left(e^{-G_0^2/\Omega_X} - e^{-G_0^2/(2\Omega_X)}e^{-G_0^2/(2\Omega_Z)}\right), \\ \hspace{6cm} \Omega_X \neq \Omega_Z \\ e^{-G_0^2/(2\Omega_X)} - \dfrac{G_0^2 + 2\Omega_X}{2\Omega_X} e^{-G_0^2/\Omega_X}, \quad \Omega_X = \Omega_Z \end{cases}$$
(39)

and

$$\Pr\left\{\sqrt{2}X < G_0 \cap U > G_0\right\}$$
$$= \begin{cases} \dfrac{\Omega_Z}{\Omega_X - \Omega_Z}\left[e^{-\frac{G_0^2}{2}\left(\frac{1}{\Omega_X}+\frac{1}{\Omega_Z}\right)} - e^{-G_0^2/\Omega_Z}\right], & \Omega_X \neq \Omega_Z \\ \dfrac{G_0^2}{2\Omega_X} e^{-G_0^2/\Omega_X}, & \Omega_X = \Omega_Z. \end{cases}$$
(40)

A closed-form expression for the AOR $N_I$ of a cooperative diversity system employing SR relaying can be obtained by applying (28), (35), and (38)–(40) in (37). We note that typically $Y_0 = G_0$ is adopted [9].

The AOD is obtained by inserting (34) and (37) into (12).

## IV. ASYMPTOTIC BEHAVIOR OF AOR AND AOD

While the analytical expressions for the AOR and AOD derived in the previous section are exact and easy to evaluate, they do not provide much insight into the impact of the various channel and system parameters on system performance and design. Thus, in this section, we provide simple high SNR approximations for the AOR and AOD of the considered cooperative diversity protocols, which reveal the influence of the rate $R_0$, the mean squared channel gains $\Omega_X, \Omega_Y, \Omega_Z$, and the Doppler frequencies $f_{mS}, f_{mR}, f_{mD}$. We note that in the subsequent analysis the condition $\Gamma_0 \to \infty$ is equivalent to the condition $G_0 \to 0$, since $R_0$ is assumed to be fixed.

### A. Direct Transmission

For high SNRs, the OP (15) simplifies to the well-known asymptotic expression [9]

$$F_X \sim \frac{2^{R_0} - 1}{\Omega_X \Gamma_0}, \quad \text{as} \quad \Gamma_0 \to \infty \tag{41}$$

where $f(x) \sim g(x)$ means that $f(x)$ and $g(x)$ are asymptotically equivalent; cf. Appendix C.

The high SNR approximation of the AOR can be obtained from (16) by using only the first term of the Maclaurin series in (C.2). This leads to

$$N_I \sim \sqrt{2\pi} f_{mX} \sqrt{\frac{2^{R_0} - 1}{\Omega_X \Gamma_0}}. \tag{42}$$

Combining (12), (41), and (42), we obtain for the AOD the high SNR approximation

$$T_X \sim \frac{1}{\sqrt{2\pi} f_{mX}} \sqrt{\frac{2^{R_0} - 1}{\Omega_X \Gamma_0}}. \tag{43}$$

### B. Variable-Gain AF Relaying

For high SNR, the OP in (18) simplifies to the known asymptotic result [9]

$$P_{\text{out}} \sim \frac{\Omega_Y + \Omega_Z}{2\Omega_X \Omega_Y \Omega_Z} \left(\frac{2^{R_0} - 1}{\Gamma_0}\right)^2. \tag{44}$$

The corresponding high SNR approximation for the AOR is provided in the following theorem.

*Theorem 4:* For high SNR, the AOR of a cooperative diversity system employing variable-gain AF relaying can be approximated as

$$N_I \sim \frac{4\sqrt{2\pi}}{3} \left( \frac{1}{\Omega_X \Omega_Z} \frac{f_{mX}^3 \Omega_X^{3/2} - f_{mZ}^3 \Omega_Z^{3/2}}{f_{mX}^2 \Omega_X - f_{mZ}^2 \Omega_Z} \right.$$
$$\left. + \frac{1}{\Omega_X \Omega_Y} \frac{f_{mX}^3 \Omega_X^{3/2} - f_{mY}^3 \Omega_Y^{3/2}}{f_{mX}^2 \Omega_X - f_{mY}^2 \Omega_Y} \right)$$
$$\times \left(\frac{2^{2R_0} - 1}{\Gamma_0}\right)^{3/2}. \tag{45}$$

*Proof:* See Appendix C.

Assuming $f_{mS} = f_{mR} = f_{mD} = f_m$, (45) simplifies to

$$N_I \sim \frac{4\sqrt{\pi} f_m}{3} \left( \frac{1}{\Omega_X \Omega_Z} \frac{\Omega_X + \sqrt{\Omega_X \Omega_Z} + \Omega_Z}{\sqrt{\Omega_X} + \sqrt{\Omega_Z}} \right.$$
$$\left. + \frac{1}{\Omega_X \Omega_Y} \frac{\Omega_X + \sqrt{\Omega_X \Omega_Y} + \Omega_Y}{\sqrt{\Omega_X} + \sqrt{\Omega_Y}} \right)$$
$$\times \left(\frac{2^{2R_0} - 1}{\Gamma_0}\right)^{3/2}. \tag{46}$$

A high SNR approximation of the AOD $T_I$ is straightforwardly obtained by inserting (44) and (45) [or (46)] into (12).

### C. DF Relaying

For high SNRs, the OP in (22) simplifies to the known asymptotic result [9, eq. (18)]

$$P_{\text{out}} \sim \frac{2^{2R_0} - 1}{\Omega_Y \Gamma_0}. \tag{47}$$

*Theorem 5:* The AOR of a cooperative diversity system employing DF relaying can be approximated as

$$N_I \sim \sqrt{2\pi} f_{mY} \sqrt{\frac{2^{2R_0} - 1}{\Omega_Y \Gamma_0}}. \tag{48}$$

*Proof:* See Appendix D.

A high SNR approximation of the AOD $T_I$ is obtained by inserting (47) and (48) into (12).

### D. Selection DF Relaying

For high SNR, (34) simplifies to the known asymptotic result [9, eq. (22)]

$$P_{\text{out}} \sim \frac{\Omega_Y + \Omega_Z}{2\Omega_X \Omega_Y \Omega_Z} \left(\frac{2^{2R_0} - 1}{\Gamma_0}\right)^2. \tag{49}$$

*Theorem 6:* The AOR of a cooperative diversity system employing SR relaying can be approximated as

$$N_I \sim \sqrt{\pi} \left( \frac{f_{mX} \sqrt{\Omega_X} + f_{mY} \sqrt{\Omega_Y/2}}{\Omega_X \Omega_Y} \right.$$
$$\left. + \frac{2\sqrt{2}}{3\Omega_X \Omega_Z} \frac{f_{mZ}^3 \Omega_Z^{3/2} - f_{mX}^3 \Omega_X^{3/2}}{f_{mZ}^2 \Omega_Z - f_{mX}^2 \Omega_X} \right)$$
$$\times \left(\frac{2^{2R_0} - 1}{\Gamma_0}\right)^{3/2}. \tag{50}$$

*Proof:* The theorem is proved straightforwardly by exploiting the asymptotic equivalence

$$\Pr\{\sqrt{2}X > G_0 \cap U < G_0\} + \Pr\{\sqrt{2}X < G_0 \cap U > G_0\}$$
$$\sim G_0^2/(2\Omega_X) \tag{51}$$

along with (D.1), (D.2), and (D.5).

A high SNR approximation of the AOD $T_I$ is obtained by inserting (49) and (50) into (12).

TABLE I
HIGH SNR APPROXIMATIONS OF AOR AND AOD FOR A SYMMETRIC COOPERATIVE DIVERSITY NETWORK UTILIZING THE AF, DF, AND SR PROTOCOLS, RESPECTIVELY. $\bar{\gamma} = \Omega \Gamma_0$ IS THE AVERAGE RECEIVED SNR AT THE THREE MOBILE NODES THAT INTRODUCE THE SAME MAXIMUM DOPPLER FREQUENCY $f_m$, I.E., $f_{mS} = f_{mR} = f_{mD} = f_m$ AND $\Omega_X = \Omega_Y = \Omega_Z = \Omega$. FOR COMPARISON, WE ALSO PROVIDE RESULTS FOR DIRECT TRANSMISSION WITH A SINGLE RECEIVE ANTENNA (DIRECT) AND TWO RECEIVE ANTENNAS ($1 \times 2$ SIMO)

| Protocol | OP | AOR | AOD |
|---|---|---|---|
| Direct | $\dfrac{2^{R_0} - 1}{\bar{\gamma}}$ | $\dfrac{2\sqrt{\pi} f_m \sqrt{2^{R_0} - 1}}{\sqrt{\bar{\gamma}}}$ | $\dfrac{\sqrt{2^{R_0} - 1}}{2\sqrt{\pi} f_m \sqrt{\bar{\gamma}}}$ |
| $1 \times 2$ SIMO | $\dfrac{(2^{R_0} - 1)^2}{2\bar{\gamma}^2}$ | $\dfrac{2\sqrt{\pi} f_m (2^{R_0} - 1)^{3/2}}{\bar{\gamma}^{3/2}}$ | $\dfrac{\sqrt{2^{R_0} - 1}}{4\sqrt{\pi} f_m \sqrt{\bar{\gamma}}}$ |
| AF | $\dfrac{(2^{2R_0} - 1)^2}{\bar{\gamma}^2}$ | $\dfrac{4\sqrt{\pi} f_m (2^{2R_0} - 1)^{3/2}}{\bar{\gamma}^{3/2}}$ | $\dfrac{\sqrt{2^{2R_0} - 1}}{4\sqrt{\pi} f_m \sqrt{\bar{\gamma}}}$ |
| DF | $\dfrac{2^{2R_0} - 1}{\bar{\gamma}}$ | $\dfrac{2\sqrt{\pi} f_m \sqrt{2^{2R_0} - 1}}{\sqrt{\bar{\gamma}}}$ | $\dfrac{\sqrt{2^{2R_0} - 1}}{2\sqrt{\pi} f_m \sqrt{\bar{\gamma}}}$ |
| SR | $\dfrac{(2^{2R_0} - 1)}{\bar{\gamma}^2}$ | $\dfrac{(\sqrt{2} + 3)\sqrt{\pi} f_m (2^{2R_0} - 1)^{3/2}}{\bar{\gamma}^{3/2}}$ | $\dfrac{\sqrt{2^{2R_0} - 1}}{(\sqrt{2} + 3)\sqrt{\pi} f_m \sqrt{\bar{\gamma}}}$ |

Assuming $f_{mS} = f_{mR} = f_{mD} = f_m$, (50) simplifies to

$$N_I \sim \sqrt{\pi} f_m \left[ \frac{\sqrt{2}}{\sqrt{\Omega_X} \Omega_Y} + \frac{1}{\Omega_X \sqrt{\Omega_Y}} \right.$$
$$\left. + \frac{4}{3\Omega_X \Omega_Z} \frac{\Omega_X + \sqrt{\Omega_X \Omega_Z} + \Omega_Z}{\sqrt{\Omega_X} + \sqrt{\Omega_Z}} \right]$$
$$\times \left( \frac{2^{2R_0} - 1}{\Gamma_0} \right)^{3/2}. \qquad (52)$$

A summary of the OPs, AORs, and AODs of symmetric networks with $f_{mS} = f_{mR} = f_{mD} = f_m$ and $\Omega_X = \Omega_Y = \Omega_Z = \Omega$ is provided in Table I. Besides the results for the three considered cooperative diversity protocols, Table I also includes the results for direct transmission with one and two receive antennas [22]. In the latter case, we have a $1 \times 2$ single-input–multiple-output (SIMO) system with independent identically distributed (i.i.d.) Rayleigh fading branches and maximal ratio combining.

### E. Diversity Gain

Based on the high SNR approximations derived in Sections IV-A–IV-D, the asymptotic AOR and AOD of the considered cooperative diversity protocols can be expressed as

$$N_I \sim n_k (\bar{f}_m, \bar{\Omega}) \left( \frac{2^{2R_0} - 1}{\Gamma_0} \right)^{d_k - 1/2}, \quad \text{as } \Gamma_0 \to \infty \qquad (53)$$

$$T_I \sim t_k (\bar{f}_m, \bar{\Omega}) \left( \frac{2^{2R_0} - 1}{\Gamma_0} \right)^{1/2}, \quad \text{as } \Gamma_0 \to \infty \qquad (54)$$

where $d_k$ ($k =$ AF, DF, or SR) denotes the diversity gain of the protocol, i.e., $d_{\text{AF}} = 2$, $d_{\text{DF}} = 1$, and $d_{\text{SR}} = 2$ [9]. For direct transmission, $d_{\text{DT}} = 1$ is valid. The SNR-independent functions $n_k(\cdot)$ and $t_k(\cdot)$ in (53) and (54), respectively, can be determined straightforwardly for each of the three protocols from (44), (45), and (47)–(50). These functions depend on the maximum Doppler frequencies of the mobile nodes $\bar{f}_m = (f_{mX}, f_{mY}, f_{mZ})$ and the mean squared channel gains $\bar{\Omega} = (\Omega_X, \Omega_Y, \Omega_Z)$.

For high SNR and on a double-logarithmic scale, both the AOR and the AOD decay linearly with the SNR $\Gamma_0$; cf. (53) and (54). Thereby, the asymptotic slopes of the AOR and AOD curves are given by $-(d_k - 1/2)$ and $-1/2$, respectively. Thus, similar to the OP, whose asymptotic slope is $-d_k$[9], the AOR strongly benefits from an increased diversity order $d_k$. Therefore, the OP and the AOR of the AF and SR protocols decay much faster with increasing SNR than that of the DF protocol and direct transmission. In contrast, the asymptotic slope of the AOD curves is independent of the diversity order and the AOD curves for all considered cooperative diversity protocols and direct transmission are parallel for high SNR. In other words, the large decrease in OP with increasing SNR achieved by the AF and SR protocols compared to the DF protocol and direct transmission is mainly due to a decrease in the frequency of outage events (which is manifested in the AOR) as opposed to a decrease in the duration of individual outage events (which is manifested in the AOD).

We note that the fact that the AOD does not benefit from the diversity gain $d_k$ is not limited to cooperative diversity systems but also applies to conventional SIMO systems with maximal ratio combining; cf., Table I.

### F. Outage Rate Versus Outage Probability

Based on (53) and (54), we can establish the following high SNR relationships between the OP and the AOR and AOD

$$N_I \sim h_k (\bar{f}_m, \bar{\Omega}) (P_{\text{out}})^{\frac{d_k + 1}{4}}, \quad \text{as } \Gamma_0 \to \infty \qquad (55)$$

$$T_I \sim \frac{1}{h_k (\bar{f}_m, \bar{\Omega})} (P_{\text{out}})^{\frac{3 - d_k}{4}}, \quad \text{as } \Gamma_0 \to \infty \qquad (56)$$

where the function $h_k(\cdot)$ can be determined straightforwardly for each of the three considered protocols by combining (44), (45), and (47)–(50), respectively.

Interestingly, while the AOR decays faster with decreasing $P_{\text{out}}$ if the diversity order $d_k$ is increased from one to two, the opposite is true for the AOD. Thus, if a certain target $P_{\text{out}}$ is required and can be achieved with different system architectures (at different SNRs), an architecture with $d_k = 1$ (DF, direct transmission) will lead to a larger AOR and a smaller AOD than an architecture with $d_k = 2$ (AF, SR). In systems with strict delay constraints, more frequent outages of shorter durations may be preferable, whereas in systems that conserve energy by completely switching off the receiver, less frequent, longer outages may be preferable to avoid frequent switching between the on and off modes.

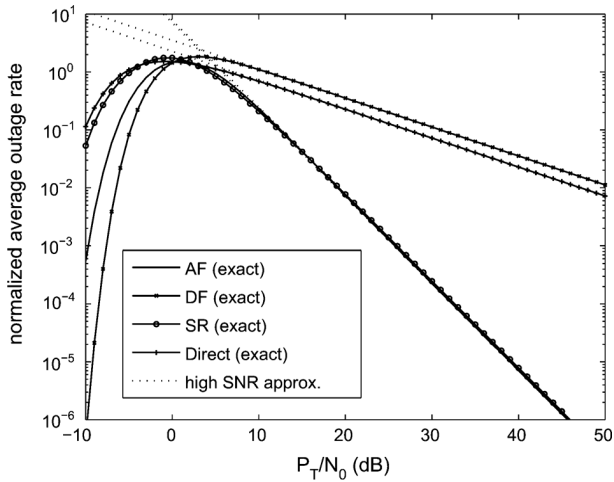

Fig. 1. Normalized AOR versus transmit SNR of symmetric cooperative diversity systems with AF, DF, and SR relaying and direct transmission ($\Omega_X = \Omega_Y = \Omega_Z = 1$).

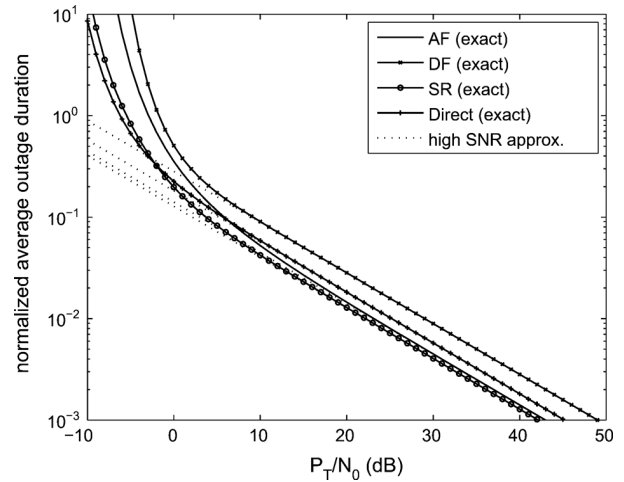

Fig. 2. Normalized AOD versus transmit SNR of symmetric cooperative diversity systems with AF, DF, and SR relaying and direct transmission ($\Omega_X = \Omega_Y = \Omega_Z = 1$).

## V. NUMERICAL RESULTS AND DISCUSSION

In this section, we study the AOR and AOD of some example networks using AF, DF, and SR relaying. To this end, we evaluate the analytical expressions developed in Sections III and IV, respectively. All results shown in this section have been validated by computer simulations. We omit the simulation results here for clarity of presentation. Throughout this section, we assume $f_{mS} = f_{mR} = f_{mD} = f_m$ and show results for the normalized AOR ($N_I/f_m$) and normalized AOD ($T_I f_m$), respectively. Furthermore, we set the target rate to $R_0 = 0.5$ b/s/Hz, i.e., the system operates in the low-spectral efficiency regime, where cooperative diversity has been shown to have significant performance benefits [9].

In Figs. 1 and 2, we show, respectively, the normalized AOR and AOD versus the transmit SNR $\Gamma_0 = P_T/N_0$, for the considered diversity protocols and direct transmission. A symmetric network with $\Omega_X = \Omega_Y = \Omega_Z = \Omega$ is assumed. Besides the exact AORs and AODs, we also show the asymptotic approximations derived in Section IV.

Figs. 1 and 2 confirm the tightness of these approximations for sufficiently high SNR. As predicted in Section IV-E, the diversity gain of the AF and SR protocols is reflected in the AOR but not in the AOD. Nevertheless, the AF and SR protocols still achieve a lower AOD than the DF protocol and direct transmission for high SNR. As an example, we consider a system with $f_m T = 10^{-3}$ and an SNR of 20 dB. In this case, Fig. 1 suggests that on average for the SR protocol, the AF protocol, the DF protocol, and direct transmission an outage event happens every $1/(N_I T) = 1.3 \times 10^5, 1.4 \times 10^5, 2.8 \times 10^3$, and $4.3 \times 10^3$ coding blocks, respectively. At the same time, Fig. 2 shows that on average these outage events last $T_I/T = 13, 14, 28$, and 18 coding blocks, respectively. This example nicely illustrates that while the differences in the AOD between the different cooperative diversity schemes are relatively small, the differences in the AOR are major, i.e., as mentioned in Section IV, the large OP gains achievable with the SR and AF protocols compared to direct transmission are mainly due to a decrease in the frequency of outage events rather than a decrease in the duration of individual outage events. Furthermore, if, for example, the considered cooperative diversity systems employ ARQ, the results in Fig. 2 show that the interval between retransmissions should be at least 13, 14, and 28 coding blocks for the SR, AF, and DF protocols, respectively.

Next, we consider a network with a comparatively strong direct link (scenario 1) and a network with a comparatively weak direct link (scenario 2). The corresponding AORs and AODs are shown in Figs. 3 and 4, respectively. While the strength of the direct link does not have any influence on the asymptotic slope of the AOR and AOD curves, it does affect the relative performance. For example, for a strong direct link, direct transmission achieves practically the same AOD as SR and AF relaying, i.e., in this case, the direct link dominates the performance of these relaying protocols. In contrast, in case of a weak direct link, all considered cooperative diversity protocols achieve a substantially lower AOD than direct transmission, as intuitively expected. Interestingly, at high SNR, both the AOR and the AOD of DF relaying are not significantly affected by the strength of the direct link, i.e., the performance limiting factor of DF relaying is the relayed link. It is interesting to connect the AOR and AOD to the coherence time of the channel $T_{\text{coh}}$. Assuming $T_{\text{coh}} \approx 1/f_m$, we have $N_I/f_m \approx N_I T_{\text{coh}}$, which denotes the average number of capacity outage events that occur within a single channel coherence time, or equivalently $f_m/N_I \approx 1/(N_I T_{\text{coh}})$, which denotes the average separation between two consecutive capacity outage events expressed in terms of the number of channel coherence times. Similarly, the normalized AOD $T_I f_m \approx T_I/T_{\text{coh}}$ denotes the average duration of an outage event in terms of the coherence time of the channel. For example, we observe from Fig. 3 that for SR relaying and an SNR of 20 dB, an outage event occurs on average every $1/(N_I T_{\text{coh}}) = 8.3$ and 30 coherence times if the direct link is strong and weak, respectively. Fig. 4 shows that the duration of these outage events is $T_I/T_{\text{coh}} = 4 \times 10^{-3}$ and $2 \times 10^{-2}$ coherence times for the strong and the weak direct link, respectively.

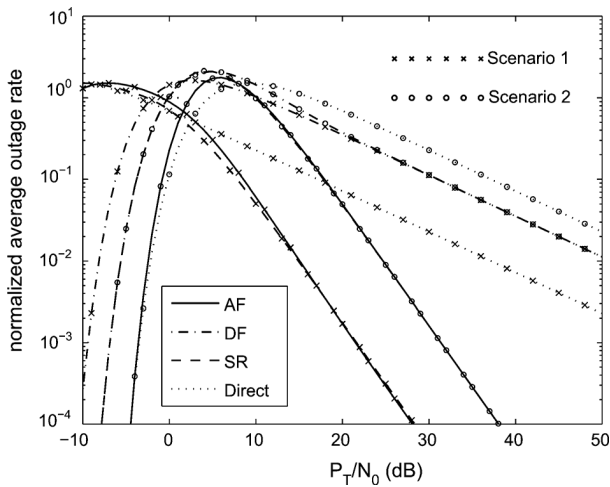

Fig. 3. Normalized AOR of asymmetric cooperative diversity systems with AF, DF, and SR relaying and direct transmission. Scenario 1 (strong direct link): $\Omega_X = 10, \Omega_Y = 1$, and $\Omega_Z = 1$. Scenario 2 (weak direct link): $\Omega_X = 0.1$, $\Omega_Y = 1$, and $\Omega_Z = 1$.

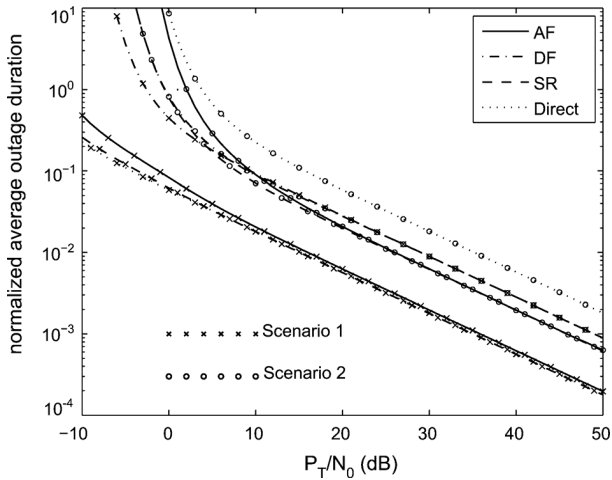

Fig. 4. Normalized AOD of asymmetric cooperative diversity systems with AF, DF, and SR relaying and direct transmission. Scenario 1 (strong direct link): $\Omega_X = 10, \Omega_Y = 1$, and $\Omega_Z = 1$. Scenario 2 (weak direct link): $\Omega_X = 0.1$, $\Omega_Y = 1$, and $\Omega_Z = 1$.

Finally, in Fig. 5, we investigate the relationship between the AOR and the OP; cf. Section IV-F. As expected from (55), for sufficiently small OPs, on the double logarithmic scale of Fig. 5, the AOR becomes a straight line with slope $(d_k+1)/4$. In other words, for small (practical) OPs, (55) and (56) can be used to quickly estimate the AOR and AOD, respectively, from the OP. For example, Fig. 5 shows that for an OP of $10^{-5}$, an outage event occurs roughly every 1000 and 100 channel coherence times if SR (AF) relaying and DF relaying (direct transmission) are used, respectively. Correspondingly, the duration of these outage events is roughly $10^{-2}$ and $10^{-3}$ channel coherence times for SR (AF) relaying and DF relaying (direct transmission), respectively.

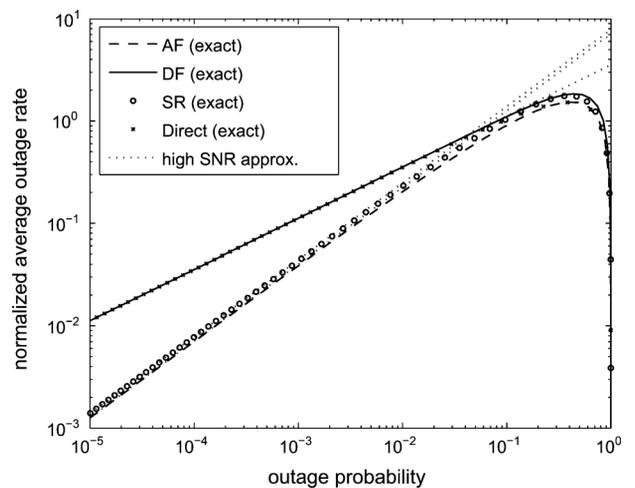

Fig. 5. AOR versus OP of asymmetric cooperative diversity systems with AF, DF, and SR relaying and direct transmission ($\Omega_X = \Omega_Y = \Omega_Z$).

## VI. CONCLUSION

In this paper, we have analyzed the AOR and the AOD of cooperative diversity systems employing AF, DF, and SR relaying in Rayleigh fading channels with mobile nodes. In contrast to the OP, the AOR and the AOD provide information about the temporal correlation of capacity outage events in slowly time-varying channels. Besides exact analytical expressions, we also developed asymptotically tight high SNR approximations for the AOR and AOD, which provide significant insight into the influence of the Doppler frequencies of the nodes, the relative strength of the involved links, the SNR, and the target transmission rate. In particular, we show that for high SNR and a double logarithmic scale, both the AOR and the AOD depend linearly on the SNR. However, while the slope of the AOR curves is affected by the diversity gain of the channel in a similar manner as the OP, the slope of the AOD curves is equal to $-1/2$ independent of the adopted cooperative protocol.

The derived AOR and AOD expressions are useful for all system design problems that are influenced not only by the OP itself but also by the frequency and duration of outage events. Such design problems include the dimensioning of the retransmission interval of ARQ systems, the scheduling slot duration of multiuser systems, and the duration of the sleep mode of energy saving receivers.

## APPENDIX A
## PROOFS OF THEOREMS 1 AND 2

Based on (17) the OP of AF relaying can be expressed as

$$F_G(G_0) = \Pr\left\{\sqrt{X^2(t) + \frac{Y^2(t)Z^2(t)}{Y^2(t)+Z^2(t)+C_1}} \leq G_0\right\}$$

$$= \int_0^{G_0} \Pr\left\{\frac{Y^2 Z^2}{Y^2+Z^2+C_1} \leq G_0^2 - x^2\right\} f_X(x)dx$$

(A.1)

where $f_X(x)$ is given in (1) and the cdf of random variable $Y^2Z^2/(Y^2+Z^2+C_1)$ at threshold $G_0^2-x^2$ is given by [28, eq. (14)]

$$\Pr\left\{\frac{Y^2Z^2}{Y^2+Z^2+C_1} \le G_0^2 - x^2\right\}$$
$$= 1 - 2\sqrt{\frac{(G_0^2-x^2)(G_0^2-x^2+C_1)}{\Omega_Y\Omega_Z}}$$
$$\times \exp\left(-\frac{\Omega_Y+\Omega_Z}{\Omega_Y\Omega_Z}(G_0^2-x^2)\right)$$
$$\times K_1\left(2\sqrt{\frac{(G_0^2-x^2)(G_0^2-x^2+C_1)}{\Omega_Y\Omega_Z}}\right). \quad (A.2)$$

Substitution of (A.2) into (A.1) and applying the change of variables $a = G_0^2 - x^2$ yields (18), thus completing the proof of Theorem 1.

In order to determine the LCR of $G(t)$ in (17), we first find its time derivative

$$\dot{G} = \frac{X}{G}\dot{X} + \frac{Z^2(Z^2+C_1)}{(Y^2+Z^2+C_1)^2}\frac{Y}{G}\dot{Y} + \frac{Y^2(Y^2+C_1)}{(Y^2+Z^2+C_1)^2}\frac{Z}{G}\dot{Z}. \quad (A.3)$$

Conditioned on $Y=y$ and $Z=z$, the joint pdf $f_{G\dot{G}}(g,\dot{g})$ is given by

$$f_{G\dot{G}}(g,\dot{g}) = \int_{y=0}^{\infty}\int_{z=0}^{\infty} f_{G\dot{G}|YZ}(g,\dot{g}|y,z) f_Y(y) f_Z(z)\,dy\,dz \quad (A.4)$$

where $f_Y(y)$ and $f_Z(z)$ are the pdfs of the independent channel gains $Y$ and $Z$ of the relayed path, respectively. The conditional joint pdf $f_{G\dot{G}|YZ}(g,\dot{g}|y,z)$ can be expressed as

$$f_{G\dot{G}|YZ}(g,\dot{g}|y,z) = f_{\dot{G}|GYZ}(\dot{g}|g,y,z) f_{G|YZ}(g|y,z) \quad (A.5)$$

where $f_{G|YZ}(g|y,z)$ is the conditional pdf of $G$, given $Y=y$ and $Z=z$. Applying a simple random variable (RV) transformation, we obtain

$$f_{G|YZ}(g|y,z) = \frac{2g}{\Omega_X}\exp\left[-\frac{1}{\Omega_X}\left(g^2 - \frac{y^2z^2}{y^2+z^2+C_1}\right)\right],$$
$$\text{for } 0 \le \frac{yz}{\sqrt{y^2+z^2+C_1}} \le g. \quad (A.6)$$

In (A.5), $f_{\dot{G}|GYZ}(\dot{g}|g,y,z)$ is the conditional pdf of $\dot{G}$, given $G=g$, $Y=y$, and $Z=z$. Hence, $\dot{G}$ is a linear combination of three independent zero-mean Gaussian RVs, $\dot{X}$, $\dot{Y}$, and $\dot{Z}$, with variances given by (3). Thus, $\dot{G}$ is also a Gaussian RV with zero mean and variance

$$\sigma_{\dot{G}|GYZ}^2 = \left(g^2 - \frac{y^2z^2}{y^2+z^2+C_1}\right)\frac{\sigma_{\dot{X}}^2}{g^2}$$
$$+ \frac{y^2z^4(z^2+C_1)^2}{(y^2+z^2+C_1)^4}\frac{\sigma_{\dot{Y}}^2}{g^2} + \frac{z^2y^4(y^2+C_1)^2}{(y^2+z^2+C_1)^4}\frac{\sigma_{\dot{Z}}^2}{g^2}. \quad (A.7)$$

Exploiting (A.4) and (A.5) in (10), and changing the order of integration, we obtain

$$N_G(G_0) = \int_{y=0}^{\infty}\int_{z=0}^{\infty}\left(\int_0^{\infty}\dot{g}f_{\dot{G}|GYZ}(\dot{g}|G_0,y,z)\,d\dot{g}\right)$$
$$\times f_{G|YZ}(G_0|y,z) f_Y(y) f_Z(z)\,dy\,dz \quad (A.8)$$

where the innermost integral is computed as

$$\int_0^{\infty}\dot{g}f_{\dot{G}|GYZ}(\dot{g}|G_0,y,z)\,d\dot{g} = \sqrt{\frac{\sigma_{\dot{G}|GYZ}^2}{2\pi}}. \quad (A.9)$$

Now, using (1), (A.6), (A.7), and (A.9) in (A.8), and introducing the change of variables $a = y^2z^2/(y^2+z^2+C_1)$, we obtain

$$N_G(G_0)$$
$$= \frac{4}{\sqrt{2\pi}}\frac{1}{\Omega_X\Omega_Y\Omega_Z}\exp\left(-\frac{G_0^2}{\Omega_X}\right)$$
$$\times \int_0^{G_0^2}da\int_a^{\infty}dy\left[(G_0^2-a)\sigma_{\dot{X}}^2 + \frac{a^2(a+C_1)^2}{y^2(y^2+C_1)^2}\sigma_{\dot{Y}}^2\right.$$
$$\left.+ \frac{a(y^2-a)^3}{y^4(y^2+C_1)^2}\sigma_{\dot{Z}}^2\right]^{1/2}\frac{y^3(y^2+C_1)}{(y^2-a)^2}$$
$$\times \exp\left[-\left(\frac{y^2}{\Omega_Y} + \frac{1}{\Omega_Z}\frac{a(y^2+C_1)}{y^2-a} - \frac{a}{\Omega_X}\right)\right]. \quad (A.10)$$

Equation (19) is obtained after applying another change of variables $t = 1/(y^2-a)$, thus completing the proof of Theorem 2.

## APPENDIX B
## PROOF OF THEOREM 3

In order to determine the LCR of random process $U(t)$, we first note that based on (21) the time derivative of $U(t)$ is given by

$$\dot{U} = \frac{X}{U}\dot{X} + \frac{Z}{U}\dot{Z}. \quad (B.1)$$

Conditioned on $Z=z$, the joint pdf $f_{U\dot{U}}(u,\dot{u})$ is determined as

$$f_{U\dot{U}}(u,\dot{u}) = \int_0^{\infty} f_{U\dot{U}|Z}(u,\dot{u}|z) f_Z(z)\,dz \quad (B.2)$$

where $f_Z(z)$ is the pdf of channel gain $Z$. The conditional joint pdf $f_{U\dot{U}|Z}(u,\dot{u}|z)$ can be expressed as

$$f_{U\dot{U}|Z}(u,\dot{u}|z) = f_{\dot{U}|UZ}(\dot{u}|u,z) f_{U|Z}(u|z) \quad (B.3)$$

where $f_{U|Z}(u|z)$ is the conditional pdf of $U$ for a given $Z=z$. This conditional pdf is given by

$$f_{U|Z}(u|z) = \frac{2u}{\Omega_X}\exp\left(-\frac{u^2-z^2}{\Omega_X}\right), \qquad 0 \le z \le u. \quad (B.4)$$

In (B.3), $f_{\dot{U}|UZ}(\dot{u}|u,z)$ is the conditional pdf of $\dot{U}$, given $U=u$ and $Z=z$. Because of the conditioning, $\dot{U}$ is a Gaussian RV with zero mean and variance

$$\sigma^2_{\dot{U}|UZ} = \frac{u^2-z^2}{u^2}\sigma^2_{\dot{X}} + \frac{z^2}{u^2}\sigma^2_{\dot{Z}}. \quad (B.5)$$

Introducing (B.2) and (B.3) into definition (10), and changing the order of integration, we obtain

$$N_U(G_0) = \int_0^{G_0}\left(\int_0^\infty \dot{u} f_{\dot{U}|UZ}(\dot{u}|G_0,z)\,d\dot{u}\right) \\ \times f_{U|Z}(G_0|z)f_Z(z)\,dz \quad (B.6)$$

where the innermost integral yields

$$\int_0^\infty \dot{u} f_{\dot{U}|UZ}(\dot{u}|G_0,z)\,d\dot{u} = \sqrt{\frac{\sigma^2_{\dot{U}|UZ}}{2\pi}}. \quad (B.7)$$

We obtain (29) and (31) by inserting (1), (B.4), (B.5), and (B.7) into (B.6), and then integrating with respect to variable $z$. More particularly, (29) is obtained by introducing the change of variables $t = 1+z^2(\sigma^2_{\dot{Z}}-\sigma^2_{\dot{X}})/(G_0^2\sigma^2_{\dot{X}})$ in (B.6) and then applying the definition of the incomplete Gamma function $\Gamma(\cdot,\cdot)$ [27, eq. (6.5.3)]. For $\Omega_X = \Omega_Z$, (B.6) reduces to an elementary integral, directly yielding (31). This completes the proof of Theorem 3.

## APPENDIX C
## PROOF OF THEOREM 4

For the high SNR analysis (as $\Gamma_0 \to \infty$) in Appendixes C and D, we adopt the Landau small-$o$ notation which describes the asymptotic relation between two functions $f(x)$ and $g(x)$ as $x \to x_0$ [30]. Thereby, $f(x) = o[g(x)]$, as $x \to x_0$, means that $\lim_{x\to x_0} f(x)/g(x) = 0$. Two functions $f(x)$ and $g(x)$ are said to be asymptotically equivalent as $x \to x_0$, if $\lim_{x\to x_0} f(x)/g(x) = 1$. This asymptotic equivalence is denoted by $f(x) \sim g(x)$, as $x \to x_0$, or, alternatively

$$f(x) = g(x) + o[g(x)], \quad \text{as} \quad x \to x_0. \quad (C.1)$$

For example, using this notation we can express $\exp(-x)$ as

$$\exp(-x) \sim 1 - x + \frac{x^2}{2} - \ldots, \quad \text{as} \quad x \to 0 \quad (C.2)$$

which will be exploited in the following.

For the high SNR approximation of (19), we set $C_1 = 1/\Gamma_0 = 0$, which simplifies (17) to

$$G(t) = \sqrt{X^2(t) + \frac{Y^2(t)Z^2(t)}{Y^2(t)+Z^2(t)}}. \quad (C.3)$$

Using a derivation similar to that in Appendix A, the AOR is obtained as

$$N_G(G_0) \\ = \frac{2}{\sqrt{2\pi}}\frac{1}{\Omega_X\Omega_Y\Omega_Z}\exp\left(-\frac{G_0^2}{\Omega_X}\right) \\ \times \int_0^{G_0^2}\left\{\exp\left[-a\left(\frac{1}{\Omega_Y}+\frac{1}{\Omega_Z}-\frac{1}{\Omega_X}\right)\right]\right.$$

$$\times \int_0^\infty \sqrt{(G_0^2-a)\sigma^2_{\dot{X}} + \frac{a^4t^3}{(at+1)^3}\sigma^2_{\dot{Y}} + \frac{a}{(at+1)^3}\sigma^2_{\dot{Z}}} \\ \left.\times \frac{(at+1)^2}{t^2}\exp\left[-\left(\frac{a^2 t}{\Omega_Z}+\frac{1}{t\Omega_Y}\right)\right]dt\right\}da \quad (C.4)$$

where $G_0 \to 0$. After some algebraic manipulations of the inner integral in (C.4) and the change of variables $u = 1/(tG_0^2)$, (C.4) becomes

$$N_G(G_0) = \frac{2}{\sqrt{2\pi}}\frac{1}{\Omega_X\Omega_Y\Omega_Z}\exp\left(-\frac{G_0^2}{\Omega_X}\right)G_0^5 \\ \times \int_0^1 J(b)\exp\left[-G_0^2 b\left(\frac{1}{\Omega_Y}+\frac{1}{\Omega_Z}-\frac{1}{\Omega_X}\right)\right]db \quad (C.5)$$

where

$$J(b) = \int_0^\infty h(u,b)\exp\left[-G_0^2\left(\frac{b^2}{u\Omega_Z}+\frac{u}{\Omega_Y}\right)\right]du \quad (C.6)$$

$$h(u,b) = \sqrt{1+\frac{b}{u}}\sqrt{(1-b)\left(1+\frac{b}{u}\right)^3\sigma^2_{\dot{X}} + \frac{b^4\sigma^2_{\dot{Y}}}{u^3} + b\sigma^2_{\dot{Z}}} \quad (C.7)$$

and $b = a/G_0^2$. Since $G_0 \to 0$ and $0 \le b \le 1$, the exponential functions in (C.5) can be approximated by using only the first term on the right-hand side of (C.2), yielding

$$N_G(G_0) = \frac{2}{\sqrt{2\pi}}\frac{1}{\Omega_X\Omega_Y\Omega_Z}(1+o(1))G_0^5\int_0^1 J(b)\,db, \\ \text{as } G_0 \to 0. \quad (C.8)$$

The second factor in (C.7) is upper bounded by

$$h(u,b) \le \left[(1-b)\left(1+\frac{b}{u}\right)^3\sigma^2_{\dot{X}} + \left(1+\frac{b}{u}\right)^3 b\sigma^2_{\dot{Y}}\right. \\ \left. + \left(1+\frac{b}{u}\right)^3 b\sigma^2_{\dot{Z}}\right]^{1/2} \\ = \left(1+\frac{b}{u}\right)^{3/2}\sqrt{(1-b)\sigma^2_{\dot{X}}+b\sigma^2_{\dot{Y}}+b\sigma^2_{\dot{Z}}} \quad (C.9)$$

from which we conclude that $h(u,b)$ can be tightly lower and upper bounded as

$$A + \frac{B_1}{u} + \frac{C}{u^2} \le h(u,b) \le A + \frac{B_2}{u} + \frac{C}{u^2}. \quad (C.10)$$

Coefficient $A$ in (C.10) is chosen to match the behavior of $h(u,b)$ at infinity

$$A = \lim_{u\to\infty} h(u,b) = \sqrt{(1-b)\sigma^2_{\dot{X}}+b\sigma^2_{\dot{Z}}} \quad (C.11)$$

whereas coefficient $C$ is chosen to match the behavior of $h(u,b)$ at $0$

$$C = \lim_{u \to 0} \frac{h(u,b)}{\frac{1}{u^2}} = b^2 \sqrt{(1-b)\sigma_X^2 + b\,\sigma_Y^2}. \quad (C.12)$$

In order to satisfy (C.10), coefficients $B_1$ and $B_2$ can be chosen to have an appropriate positive value depending on the parameters $b, \sigma_X^2, \sigma_Y^2$, and $\sigma_Z^2$. However, these coefficients do not have to be specified in the following derivation.

Applying bounds (C.10) in the integral in (C.6), we obtain

$$J_1(b) + B_1 J_2(b) + J_3(b) \le J(b) \le J_1(b) + B_2 J_2(b) + J_3(b) \quad (C.13)$$

where $J_1(b), J_2(b),$ and $J_3(b)$ are determined in closed form as

$$J_1(b) = \int_0^\infty A \exp\left[-G_0^2\left(\frac{u}{\Omega_Y} + \frac{b^2}{u\Omega_Z}\right)\right] du$$
$$= 2A\,b\sqrt{\frac{\Omega_Y}{\Omega_Z}} K_1\left(\frac{2G_0^2 b}{\sqrt{\Omega_Y \Omega_Z}}\right) \quad (C.14)$$

$$J_2(b) = \int_0^\infty \frac{1}{u} \exp\left[-G_0^2\left(\frac{u}{\Omega_Y} + \frac{b^2}{u\Omega_Z}\right)\right] du$$
$$= 2 K_0\left(\frac{2G_0^2 b}{\sqrt{\Omega_Y \Omega_Z}}\right) \quad (C.15)$$

$$J_3(b) = \int_0^\infty \frac{C}{u^2} \exp\left[-G_0^2\left(\frac{u}{\Omega_Y} + \frac{b^2}{u\Omega_Z}\right)\right] du$$
$$= \frac{2C}{b}\sqrt{\frac{\Omega_Z}{\Omega_Y}} K_1\left(\frac{2G_0^2 b}{\sqrt{\Omega_Y \Omega_Z}}\right) \quad (C.16)$$

where $K_0(\cdot)$ and $K_1(\cdot)$ are the modified zeroth- and first-order Bessel functions of the second kind [27, eq. (9.6.2)], respectively. From [27, eq. (9.6.8)–(9.6.9)], we have

$$K_0(z) = -\ln(z) + o[\ln(z)], \quad \text{as} \quad z \to 0 \quad (C.17)$$

and

$$K_1(z) = \frac{1}{z} + o\left(\frac{1}{z}\right), \quad \text{as} \quad z \to 0 \quad (C.18)$$

and (C.14)–(C.16) can be simplified. To solve the integral in (C.8), we integrate over these simplified versions of $J_1(b), J_2(b),$ and $J_3(b)$, which leads to

$$\int_0^1 J_1(b)db = \frac{\Omega_Y}{G_0^2} \int_0^1 \sqrt{(1-b)\sigma_X^2 + b\sigma_Z^2}\,db + o\left(\frac{1}{G_0^2}\right)$$
$$= \frac{2\Omega_Y}{3G_0^2} \frac{\sigma_X^3 - \sigma_Z^3}{\sigma_X^2 - \sigma_Z^2} + o\left(\frac{1}{G_0^2}\right) \quad (C.19)$$

$$\int_0^1 J_2(b)db = -2\ln\left(\frac{2G_0^2}{\sqrt{\Omega_Y \Omega_Z}}\right) - 2\int_0^1 \ln(b)db + o\left[\ln(G_0^2)\right]$$
$$= -2\ln\left(\frac{2G_0^2}{e\sqrt{\Omega_Y \Omega_Z}}\right) + o\left[\ln(G_0^2)\right] \quad (C.20)$$

$$\int_0^1 J_3(b)db = \frac{\Omega_Z}{G_0^2} \int_0^1 \sqrt{(1-b)\sigma_X^2 + b\sigma_Y^2}\,db$$
$$= \frac{2\Omega_Z}{3G_0^2} \frac{\sigma_X^3 - \sigma_Y^3}{\sigma_X^2 - \sigma_Y^2} + o\left(\frac{1}{G_0^2}\right) \quad (C.21)$$

as $G_0 \to 0$. Using $\lim_{z \to 0}(z\ln(z)) = 0$, (C.20) is reduced to

$$\int_0^1 J_2(b)db = o\left(\frac{1}{G_0^2}\right), \quad \text{as } G_0^2 \to 0. \quad (C.22)$$

Thus, the lower and the upper bounds in (C.13) converge to $J(b) = J_1(b) + J_3(b)$ as $G_0 \to 0$. Hence, (C.8) is finally simplified to

$$N_G(G_0) = \frac{2}{\sqrt{2\pi}} \frac{G_0^5}{\Omega_X \Omega_Y \Omega_Z}$$
$$\times \left(\int_0^1 J_1(b)\,db + \int_0^1 J_3(b)\,db\right) + o(G_0^3)$$
$$= \frac{4G_0^3}{3\sqrt{2\pi}}\left(\frac{1}{\Omega_X \Omega_Z} \frac{\sigma_X^3 - \sigma_Z^3}{\sigma_X^2 - \sigma_Z^2} + \frac{1}{\Omega_X \Omega_Y} \frac{\sigma_X^3 - \sigma_Y^3}{\sigma_X^2 - \sigma_Y^2}\right)$$
$$+ o(G_0^3), \quad \text{as} \quad G_0 \to 0. \quad (C.23)$$

Applying (3) in (C.23) yields (45), thus completing the proof of Theorem 4.

## APPENDIX D
## PROOF OF THEOREM 5

To arrive at a high SNR approximation for (27), we assume $G_0 \to 0$. Furthermore, we replace the exponential functions appearing (23), (24), (28), and (29), by the first two terms of the series expansion in (C.2). Thus, (23) simplifies to $\Pr\{U > G_0\} = 1 + o(1)$, whereas (24) yields

$$\Pr\{Y > G_0\} = 1 - \frac{G_0^2}{\Omega_Y} + o(G_0^2). \quad (D.1)$$

To approximate (28), it is sufficient to use only the first term of (C.2) yielding

$$N_Y(G_0) = \sqrt{\frac{2\sigma_Y^2}{\pi}} \frac{G_0}{\Omega_Y} + o(G_0). \quad (D.2)$$

To approximate $N_U(G_0)$ in (29), we utilize the power series expansion of the incomplete Gamma function [27, eq. (6.5.3), (6.5.4), and (6.5.29)]

$$\Gamma\left(\frac{3}{2}, x\right) = \Gamma\left(\frac{3}{2}\right) - \frac{2x^{3/2}}{3} - \sum_{k=1}^\infty \frac{(-1)^k x^{k+3/2}}{(k+3/2)k!} \quad (D.3)$$

yielding

$$\lim_{x \to 0} \frac{\Gamma\left(\frac{3}{2}, x\right) - \Gamma\left(\frac{3}{2}, \frac{\sigma_Z^2}{\sigma_X^2}x\right)}{x^{3/2}} = \frac{2}{3}\left(\frac{\sigma_Z^3}{\sigma_X^3} - 1\right) \quad (D.4)$$

where $x = W(G_0)$; cf. (30). Using (D.4) and (C.2) in (29) yields

$$N_U(G_0) = \frac{2}{3}\sqrt{\frac{2}{\pi}} \frac{G_0^3}{\Omega_X \Omega_Z} \frac{\sigma_Z^3 - \sigma_X^3}{\sigma_Z^2 - \sigma_X^2} + o(G_0^3). \quad (D.5)$$

Applying (D.1), (D.2), and (D.5) in (27), we obtain

$$N_G(G_0) = \sqrt{\frac{2\sigma_Y^2}{\pi} \frac{G_0}{\Omega_Y}} + o(G_0). \quad (D.6)$$

Combining (3) and (D.6) yields (48), which completes the proof of Theorem 5.

## ACKNOWLEDGMENT

The authors would like to thank the editor and both reviewers for their valuable suggestions and detailed comments on the first version of this paper.

**Nikola Zlatanov** (S'06) received the Dipl.Ing. and M.S. degrees in electrical engineering from Ss. Cyril and Methodius University, Skopje, Macedonia, in 2007 and 2010, respectively. Currently, he is working towards the Ph.D. degree at the University of British Columbia (UBC), Vancouver, BC, Canada.

His current research interests include the general field of wireless communications, with emphasis on cooperative diversity systems, buffer-aided relaying, and physical layer security.

Mr. Zlatanov received the Four Year Doctoral Fellowship by UBC in 2010. In 2011, he received the Kashmir Singh Manhas Scholarship in Applied Science, the Izaak Walton Killam Memorial Doctoral Scholarship, and was awarded young scientist of the year 2010 by the President of the Republic of Macedonia.

**Zoran Hadzi-Velkov** (M'97–SM'11) received the Dipl. Ing. (honors), M.Sc. (honors), and Ph.D. degrees in electrical engineering from the Ss. Cyril and Methodius University, Skopje, Macedonia, in 1996, 2000, and 2003, respectively.

In 1996, he joined the Faculty of Electrical Engineering and Information Technologies, Ss. Cyril and Methodius University, where he is currently an Associate Professor. During 2001 and 2002, on leave from the Ss. Cyril and Methodius University, he was with the International Business Machines (IBM) T.J. Watson Research Center, Yorktown Heights, NY, where he was engaged in optical communications. He has also been a Visiting Scientist at several European universities. His current research interests are in the area of digital communications over fading channels with particular emphasis on the cooperative communications.

Dr. Hadzi-Velkov is a frequent reviewer for numerous IEEE journals.

**George K. Karagiannidis** (M'97–SM'04) was born in Pithagorion, Samos Island, Greece. He received the University Diploma (5 years) and the Ph.D. degree in electrical and computer engineering from the University of Patras, Patras, Greece, in 1987 and 1999, respectively.

From 2000 to 2004, he was a Senior Researcher at the Institute for Space Applications and Remote Sensing, National Observatory of Athens, Athens, Greece. In June 2004, he joined the Aristotle University of Thessaloniki, Thessaloniki, Greece, where he is currently an Associate Professor of Digital Communications Systems in the Electrical and Computer Engineering Department and Head of the Telecommunications Systems and Networks Lab. His current research interests are in the broad area of digital communications systems with emphasis on cooperative communication, adaptive modulation, multiple-input–multiple-output (MIMO) systems, optical wireless and underwater communications. He is the author or coauthor of more than 110 technical papers published in scientific journals and presented at international conferences. He is also a coauthor of three chapters in books and author of the Greek edition of a book on telecommunications systems.

Dr. Karagiannidis has been a member of Technical Program Committees for several IEEE conferences such as ICC, GLOBECOM, etc. He is a member of the editorial board of the IEEE TRANSACTIONS ON COMMUNICATIONS, Senior Editor of the IEEE COMMUNICATIONS LETTERS, and Lead Guest Editor of the special issue on "Optical Wireless Communications" of the IEEE JOURNAL ON SELECTED AREAS ON COMMUNICATIONS. He is corecipient of the Best Paper Award of the Wireless Communications Symposium (WCS) in the IEEE International Conference on Communications (ICC'07), Glasgow, U.K. He is the Chair of the IEEE COMSOC Greek Chapter.

**Robert Schober** (S'98–M'01–SM'08–F'10) was born in Neuendettelsau, Germany, in 1971. He received the Diplom (Univ.) and Ph.D. degrees in electrical engineering from the University of Erlangen-Nuermberg, Germany, in 1997 and 2000, respectively.

From May 2001 to April 2002, he was a Postdoctoral Fellow at the University of Toronto, Toronto, ON, Canada, sponsored by the German Academic Exchange Service (DAAD). Since May 2002, he has been with the University of British Columbia (UBC), Vancouver, BC, Canada, where he is now a Full Professor and Canada Research Chair (Tier II) in Wireless Communications. His research interests fall into the broad areas of communication theory, wireless communications, and statistical signal processing.

Dr. Schober received the 2002 Heinz Maier–Leibnitz Award of the German Science Foundation (DFG), the 2004 Innovations Award of the Vodafone Foundation for Research in Mobile Communications, the 2006 UBC Killam Research Prize, the 2007 Wilhelm Friedrich Bessel Research Award of the Alexander von Humboldt Foundation, and the 2008 Charles McDowell Award for Excellence in Research from UBC. In addition, he received best paper awards from the German Information Technology Society (ITG), the European Association for Signal, Speech and Image Processing (EURASIP), IEEE International Conference on Ultra-Wideband (ICUWB 2006), the International Zurich Seminar on Broadband Communications, and European Wireless 2000. He is also the Area Editor for Modulation and Signal Design for the IEEE TRANSACTIONS ON COMMUNICATIONS.